\documentclass[multphys,vecphys]{svmult}

\usepackage{makeidx}         
\usepackage{graphicx}        
\usepackage{multicol}        
\usepackage[bottom]{footmisc}

\makeindex             

\newcommand{\aj}{AJ}	
\newcommand{\apj}{ApJ}	
\newcommand{\apjs}{ApJS}	
\newcommand{\qjras}{QJRAS}	
\newcommand{\mnras}{MNRAS}	

\begin{document}

\title*{Extragalactic Star Clusters in Merging Galaxies}
\author{Gelys Trancho\inst{1}\inst{2}}
\institute{Gemini Observatory, 670 N. A'ohoku place, 96720, Hilo, Hawaii, USA
\texttt{gtrancho@gemini.edu}
\and Universidad de La Laguna, Avenida Astr\'ofisico Francisco S\'anchez s/n, 38206, La Laguna, Tenerife, Canary Island, Spain }
%
%
\maketitle

\begin{abstract}
The study of cluster populations as tracer of galaxy evolution is now
quite possible with 8 m class telescopes and modern
instrumentation. The cluster population can be used as a good tracer
of the star forming episodes undergone by the merging system. We
present two young galaxies mergers NGC3256 and NGC4038, and the
studies about the young cluster population on those system. We found
that the clusters ages are agree with the mergers  age and their
metallicities are consistent with them being the progenitors of the
old metal rich globulars in ellipticals. 
\end{abstract}

\section{Introduction}
\label{sec:1}
The first ideas about the differences in galaxies' morphology were
that those were due to differences in the initial conditions present at
the time of formation, not to any events that may have happened during
their evolution. This way, ellipticals looked like ellipticals because they formed through a fast and turbulent colapse; while spirals would go through as low dissipation process that allowed the formation of the disk. However, in the 70's \cite{1972ApJ...178..623T} and \cite{1978IAUS...77..279S} argued that a big fraction of ellipticals was formed by the merger of two spirals, rather than being ellipticals from the start.The argument was based on observational evidences, where remnants of mergers would present structures that were due to tidal interaction of disks.
Looking for other clues to prove or disprove either argument,
researchers eventually resourced to globular clusters. Globular
clusters have always been very important for the determination of the
evolutive  history of nearby galaxies. They are bright, numerous and
provide an excellent chronometer and age metallicity indicator for the
star formation burst from which they originated. In the early 90's,
these studies led to the argument that, since globular clusters are
more common in ellipticals than in spirals (higher specific
frequency), if the number of globulars is conserved during a merger,
ellipticals could no possible come from a combination of spirals
\cite{1990QJRAS..31..153V}. The question then become to determine if
globulars clusters can  be formed during and after a merger
event. Evidence soon started to build up that the answer for that is
yes (\cite{1987ApJ...320..454S}; \cite{1987nngp.proc...47B};
\cite{1993ApJ...416..576K}). Continuing along the lines of these
works, we devised as the primary goal of my thesis to investigate the
formation of elliptical galaxies as a result of the merger of two
spirals, by using the star cluster populations present in the merger
as tracers of the merger evolution through a study of their ages,
metallicities, masses, sizes, etc. For this we selected a sample of
merger systems of different ages, from very young (a few Myrs) to
``bona fide'' merger remnants (older than 700 Myr). In this work we
will present the results of the two younger mergers in our sample:
NGC4038/39 and NGC3256. 

\section{Observations}
\label{sec:2}
Globular clusters may be bright and numerous, but in the extragalactic
domain, even for the closest mergers the light collecting power of
8m-class telescopes and the high spatial resolution of HST are needed
to obtain the data required for a study like this. Starting with HST V
and I images from the HST archive, we constructed colour-magnitude
diagrams to select our cluster candidates:
\begin{itemize}
\item if the candidates are fainter than V=23, we have to use a
combination of optical and near-IR photometry, comparing the data on
colour-colour diagrams with Single Stellar Population (SSP) models. 
\item if the cluster candidates are brighter than V=23, spectroscopic
follow-up is possible, and we then compare the properties of the
cluster in age-metallicity spectral index diagrammes against the
values from SSP models.\\
Our spectroscopy data are mostly Gemini GMOS Multi-Object Spectroscopy
(MOS), with one merger observed with the GMOS Integral Field Unit
(IFU). Spectra were centered around 5000\AA  to include several Balmer
lines, MgI at 5100\AA, and a few iron features. 
\end{itemize}

\section{Data reduction and analysis}
\label{sec:3}

We followed the standard MOS data reduction procedures using the
GEMINI GMOS IRAF package. For each galaxy, the clusters naturally
separated in two samples:

\begin{itemize}
\item Clusters presenting only absorption lines (hydrogen, magnesium and iron are the most common).\\
The radial velocity is accurately determined by using cross correlation of the absorption lines against a RV stars observed with the same instrumental setup.
The age/metallicity indexes are measured using INDEXF
\cite{2001MNRAS.326..959C}, which correctly takes into account the
error propagation from random noise and from the uncertainty in the
velocity determination. Once the indexes are measured, we plot them
into a diagram such as H$\gamma$ $\times$\ [Mgfe] or H$\beta$
$\times$\ [MgFe] \cite{1998AJ....116.2206S}, comparing with indexes
from SSP models to obtain a first determination of the age and metallicity.\\
We then compare our observed spectrum with the model spectrum for that
age and metallicity, and determine the internal reddening using the
continuum shape. After correcting for the reddening, we re-measure the
indexes and refine the age and metallicity determination.

\item Clusters with the spectrum dominated by emission lines.\\ 
Those are by definition young clusters, which still contain enough hot
stars to ionize the surrounding gas. For these, we determine the radial velocity using the known emission lines in the spectrum. 
We then estimate the extinction from the gas in which the cluster is
embedded, using the H$gamma$/H$beta$ ratio, assuming a Case B
recombination and the  \cite{1984MNRAS.211..507E} extinction curve. Once
the reddening is corrected, we calculate the gas metalicity as a
surrogate for the embedded cluster, using the [O\,{\rm
III}]$\lambda\lambda$4363,4959+5007\AA\ lines and the formulae from
\cite{1992IAUS..149..497V} and \cite{1985ApJS...57..173M} for the
oxygen abundance.  
\end{itemize}

\section{Results}
\label{sec:4}

\begin{itemize}

\item NGC4038:\\
The Antennae galaxy. This system is at a distance of 19.2 Mpc, and is
a young merger of 200 Myr, with the two merging galaxies still
identifiable. From HST and GMOS imaging we selected 29 cluster
candidates, of which 16 were confirmed, one being located in the tidal
tail. Eight of had a pure absorption spectrum, four have an emission
spectrum and four are mixed. In summary, the clusters present the
following properties: 

\begin{itemize}
\item  Magnitudes $16 < V < 21.5$ (Reddening corrected)
\item The absorption line clusters span a range of ages between 70 to just over 300 Myr, with solar metalicity.
\item The emission line clusters are obviously younger, less than 10
Myr, and the metalicity obtained from the emission gas is a little under solar.
\item We find that the internal extinction can be quite large, up to
A$_V \sim$ 2.5mag, with the more reddened clusters located closer to
the nucleus of the secondary galaxy (NGC4039).
\end{itemize}

\item NGC3256\\
This is a merger system twice as far as the Antennae and slightly
older, but with the two nuclei still separated. The HST and GMOS
imaging provided 109 cluster candidates, of which only 31 were
spectroscopically confirmed (this galaxy is at lower galactic latitude
so the field was quite contaminated by foreground stars). Three are
located in the tidal tail.  

\begin{itemize}
\item  They were still quite bright, despite being twice the distance
of the Antennae ones, with $17.5 < V < 22.5$ . ( Reddening corrected)
\item  The absorption line clusters span a range of ages quite similar
to those in the Antennae, 80 to 300 Myr, but with solar or higher than
solar metallicity. 
\item The young cluster (those with emission lines) seem to present a
slightly lower metallicity. 
\item Again we measure quite a large reddening internal to the galaxy,
up to A$_V \sim$ 3.5mag, with more reddened clusters being the ones
closer to the center as expected.
\item We have also estimated the size of the clusters as between 1 to
10 pc, except the ones in the tidal tail, which are larger (10 to 18 pc). 
 \end{itemize}

 \end{itemize}

\section{Conclusion}
\label{sec:5}

The cluster population is a good tracer of the star forming episodes
undergone by a merging system. As we initially proposed, all galaxies
in our sample indicate that these new generations of clusters are
formed at different epochs and with different metallicities. The
cluster population is also a good indicator of the evolution of the
merging system, being more uniformly older for the older mergers.\\
The new populations are systematically more metal rich, which is
consistent with the ideas of merging events being what causes the
bimodal distribution of clusters in ellipticals. The overall
characteristics (sizes, masses) of the new clusters are consistent
with them being the progenitors of the old metal rich globulars in
ellipticals. \\ 
Clusters close to the centre of the merging system can be strongly
reddened. This can have a very drastic effect in the resulting colours
or in the line indexes and, if not corrected,  will yield cluster ages
much larger than the actual values.\\  
The presence of clusters in the tidal tails of the merging systems,
indicates that clusters formed in merger events can be ejected and end
up forming part of the intragroup medium. 




\printindex

\begin{thebibliography}{}
\bibitem[Burstein(1987)]{1987nngp.proc...47B} Burstein, D.\ 1987, Nearly 
Normal Galaxies.~From the Planck Time to the Present, 47 
\bibitem[Cenarro et al.(2001)]{2001MNRAS.326..959C} Cenarro, A.~J., 
Cardiel, N., Gorgas, J., Peletier, R.~F., Vazdekis, A., \& Prada, F.\ 2001, 
\mnras, 326, 959 
\bibitem[Edmunds \& Pagel(1984)]{1984MNRAS.211..507E} Edmunds, M.~G., \& 
Pagel, B.~E.~J.\ 1984, \mnras, 211, 507 
\bibitem[Kumai et al.(1993)]{1993ApJ...416..576K} Kumai, Y., Hashi, Y., \& 
Fujimoto, M.\ 1993, \apj, 416, 576 
\bibitem[Meyer(1985)]{1985ApJS...57..173M} Meyer, J.-P.\ 1985, \apjs, 57, 173 
\bibitem[Schweizer(1978)]{1978IAUS...77..279S} Schweizer, F.\ 1978, IAU 
Symp.~ 77: Structure and Properties of Nearby Galaxies, 77, 279 
\bibitem[Schweizer et al.(1987)]{1987ApJ...320..454S} Schweizer, F., Ford, 
W.~K.~J., Jederzejewski, R., \& Giovanelli, R. 1987, \aj, 320, 454 
\bibitem[Schweizer \& Seitzer(1998)]{1998AJ....116.2206S} Schweizer, F., \& 
Seitzer, P.\ 1998, \aj, 116, 2206 
\bibitem[Toomre \& Toomre(1972)]{1972ApJ...178..623T} Toomre, A., \& 
Toomre, J.\ 1972, \apj, 178, 623 
\bibitem[van den Bergh(1990)]{1990QJRAS..31..153V} van den Bergh, S.\ 1990, \qjras, 31, 153 
\bibitem[Vacca \& Conti(1992)]{1992IAUS..149..497V} Vacca, W.~D., \& Conti, 
P.~S.\ 1992, IAU Symp.~149: The Stellar Populations of Galaxies, 149, 497 
\end{thebibliography}
\end{document}